# Synthesis of sodium cobaltate Na$_x$CoO$_2$ single crystals with controlled Na ordering

I.F. Gilmutdinov[a*], I.R. Mukhamedshin[a,b], F. Rullier-Albenque[c], H. Alloul[b]

[a] *Institute of Physics, Kazan Federal University, 420008 Kazan, Russia*
[b] *Laboratoire de Physique des Solides, CNRS UMR 8502, Université Paris-Sud, 91405 Orsay, France*
[c] *Service de Physique de l'Etat Condensé, CEA Saclay, CNRS-URA2464, 91191 Gif-sur-Yvette, France*

*Corresponding author.
E-mail address: Ildar.Gilmutdinov@gmail.com (I.F. Gilmutdinov)

In this study, we synthesized single crystals of Na$_x$CoO$_2$ with $x \sim 0.8$ using the optical floating zone technique. A thorough electrochemical treatment of the samples permitted us to control the de-intercalation of Na to obtain single crystal samples of stable Na ordered phases with $x = 0.5$–$0.8$. Comparisons of the bulk magnetic properties with those observed in the Na ordered powder samples confirmed the high quality of these single crystal phases. The *ab* plane resistivity was measured for the Na ordered samples and it was quite reproducible for different sample batches. The data were analogous to those found in previous initial experimental studies on single crystals, but the lower residual resistivity and sharper anti-ferromagnetic transitions determined for our samples confirmed their higher quality.

**Introduction**

Sodium cobaltate compounds Na$_x$CoO$_2$ have attracted interest due to their diverse physical properties, such as magnetically ordered states [1], metal–insulator transition [2], and superconductivity [3]. Their high metallic conductivity accompanied by a high thermoelectric coefficient suggest promising thermoelectric applications [4]. Moreover, studies of their magnetic and transport properties have identified several distinct contributions to the thermopower. In particular, the spin entropy is enhanced by the electronic frustration associated with the triangular cobalt sublattice [5]. Further detailed investigations of strongly correlated frustrated compounds may identify key properties of thermoelectric materials to facilitate novel research directions.

The crystal structure of Na$_x$CoO$_2$ comprises Na layers intercalated between cobalt oxide layers where the Co ions form a two-dimensional triangular lattice. The mobility of the sodium ions and their electrostatic repulsion leads to the appearance of distinct phases. These phases are characterized by their sodium content and despite discrepancies in previous studies, well defined Na orderings have been clearly established for some phases [6–11]. These Na orderings were successfully achieved by systematic synthesis in the polycrystalline form based on careful selections of the stoichiometry and synthesis temperature [8,12]. A mixture of the two nearest phases appears for intermediate sodium contents. This phase purity remains a major experimental problem in polycrystalline sodium cobaltates due to the phase separation induced by the high Na ion mobility. In addition, the improper handling of samples, i.e., keeping samples in contact with humid air, leads to destruction of the phase purity and sodium content losses [13].

Studies of transport and electronic properties do require single crystals. Sodium cobaltates single crystals have been grown so far using the flux method [14,15] and the optical floating zone technique [16,17]. Sodium evaporation occurs during the crystal growth process, so 1–3% sodium excess is usually added to the target stoichiometry. However, although polycrystalline $Na_xCoO_2$ samples can be synthesized for a large range of Na contents, the single crystals can only be grown for a narrow range of large $x$ values (above 0.7). Electrochemical de-intercalation of sodium ions has been used [18,19] in order to obtain single crystals with lower Na concentrations. Analytical methods such as electron probe microanalysis, inductively coupled plasma [20], or energy dispersive X-ray were used to determine the average sodium contents of the synthesized samples.

Foo et al. investigated the bulk properties of samples where $0.3 < x < 0.75$ and established an early phase diagram for Na cobaltates [21]. They demonstrated that specific phases had magnetically ordered states or metal–insulator transitions. However, in most of the samples studied, they did not unambiguously establish the actual phase purity and sodium content of the dominant phase, which controls the physical properties. Therefore, further studies of the magnetic and transport properties of well-characterized pure phase single crystals are required to understand the effects of Na ordering on the electronic properties of the $CoO_2$ planes.

In the present study, we synthesized reproducible single crystals with pure phases and controlled Na content and ordering, which are absolute requirements for transport measurements and for determining the actual Fermi surface properties. We synthesized single crystal rods with a large Na content of $x \sim 0.8$ using an optical floating zone furnace. We then took advantage of electrochemical investigations done recently [22] on powder samples which successfully detected nine phases with narrow sodium compositions in the range of $0.5 \leq x < 1$, which agreed mostly with those identified in our previous systematic nuclear magnetic resonance (NMR) and magnetic studies [23]. Using a similar well-controlled electrochemical de-intercalation process applied to small single crystal platelets extracted from the as-grown sample rods, we demonstrate here the synthesis of Na ordered single crystals with controlled Na content. Simple observations of the (*008*) Bragg peaks by X-ray diffraction (XRD) allowed us to identify the sodium content and to establish the absence of phase mixing.

Using Na ordered single crystal samples, we demonstrate here that the *ab* plane residual resistivity obtained is much lower than that found in previous experiments. Furthermore, the temperature dependence of the resistivity exhibit singularities which are quite reproducible for different sample batches of the same phase.

**Crystal growth**

Single crystals of sodium cobaltates $Na_xCoO_2$ were grown using an optical floating zone furnace (Crystal System Inc.) equipped with four 300 W halogen lamps. First, polycrystalline samples of sodium cobaltate with the nominal composition $Na_{0.80}CoO_2$ were prepared in a solid state reaction using $Na_2CO_3$ (99.95%, Alfa Aesar) and $Co_3O_4$ (99.7%, Alfa Aesar). The stoichiometric mixture of $Na_2CO_3$ and $Co_3O_4$ was ground manually in an agate mortar for 2 h, and loaded into an alumina crucible. The mixture was sintered three times at 860°C in air. Intermediate grinding was performed in a a moisture-free atmosphere in a glove box to obtain a homogeneous composition. The total sintering time was 36 h. The crucible was then loaded into a preheated furnace to decrease the sodium losses [26]. The synthesized powder was analyzed by XRD using a Bruker D8 ADVANCE (Cu, K$\alpha_1$, and K$\alpha_2$) diffractometer. The XRD pattern obtained for one of the synthesized powders is shown in Fig. 1(a). The hexagonal space group P6$_3$/mmc (No. 194) corresponding to $Na_xCoO_2$ was used to index the XRD peaks [27]. There were no traces of unreacted components or other impurities. To prepare a polycrystalline feed rod for the crystal growth process, a rubber container was filled with the powder and hydrostatically pressed ($P$ = 55 MPa). The feed rod obtained was sintered in a vertical furnace at 1050°C for 2 h in air.

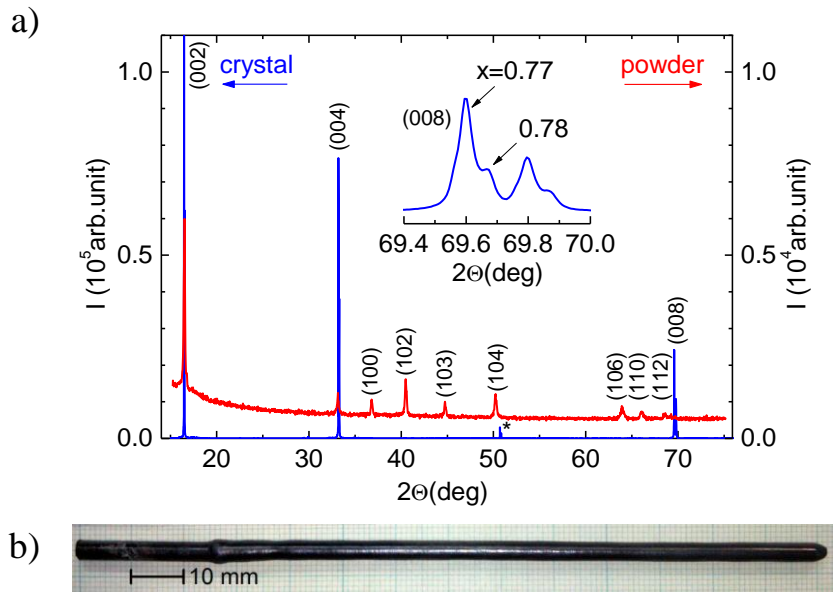

Fig. 1. (a) XRD patterns obtained for the synthesized powder $Na_{0.8}CoO_2$ feed rod (red pattern) and an as-grown crystal (blue pattern). Indices correspond to the hexagonal space group P6$_3$/mmc (No. 194). An asterisk denotes the (*006*) peak. The inset shows the enlarged (*008*) reflection of the as-grown crystal. The double peak structure of these reflections corresponds to the two Bragg peaks associated with copper K$_{\alpha1}$ and K$_{\alpha2}$ radiations. (b) Photograph of the as-grown crystal ingot. The crystallographic *c* axis is perpendicular to the main axis of the rod.

Part of the ceramic feed rod was used as a seed in the first crystal growth attempt. During the subsequent growth attempts, the previously grown single crystals were used as seeds. An oxygen-rich high pressure atmosphere ($P$ = 5 bar) was used to prevent Na losses and the formation of $Co_3O_4$ impurities. The crystal growth speed was varied in the range of 2–4 mm/h. The feed and seed rods were rotated at a speed of 30 rpm in opposite directions. The crystals were obtained in the form of cylindrical rods with an average diameter of about 5–6 mm. One of the as-grown crystals is shown in Fig. 1(b).

Single domain plates (lamella) with shiny flat surfaces were easily cleaved from the as-grown crystal. The XRD pattern obtained for a single crystal plate is also shown in Fig. 1(a), which indicates that only (*hkl*) peaks with zero *h* and *k* Miller indices were present. These peaks correspond to the diffraction from *ab* planes and they may be used to estimate the crystal lattice *c*-parameter with Bragg's law. The low background signal and the absence of other peaks confirmed the good quality of the crystals.

According to Bragg's law, the (*008*) peak with the largest *2θ* angle gives the highest resolution. The enlarged view of the (*008*) peaks for the as-grown crystal shown in the inset in Fig. 1(a) indicates splitting, thereby demonstrating that the sample comprised two phases with slightly different sodium contents. The $x = 0.77$ phase was characterized by its transition to an A type antiferromagnetic (AF) ordered state with $T_N = 22$ K [9]. The x = 0.78 phase had a magnetic transition at 9 K [28]. The electrochemical method could be used precisely tune the Na content and obtain single phase crystals [29].

**Electrochemical synthesis**

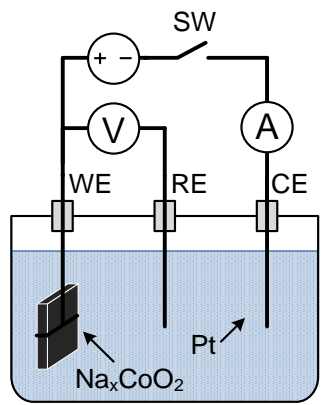

Fig. 2. Schematic showing the experimental setup for the airtight three-electrode electrochemical cell. The $Na_xCoO_2$ crystal was mounted as the working electrode and connected to the potentiostat, which is shown outside the cell (WE: working electrode, CE: counter electrode, RE: reference electrode, and SW: switch).

Electrochemical de-intercalation of sodium was used in order to obtain single crystals of sodium cobaltate $Na_xCoO_2$ with $x < 0.8$. The reduction of the Na content was conducted in an airtight electrochemical cell with three Pt wire electrodes (with a diameter of 0.5 mm and separated by distances of about 2–3 mm). A schematic of the experimental setup is shown in Fig. 2. The cell was filled with 1–2 mL of 1 M sodium perchlorate $NaClO_4$ solution (98%, Alfa Aesar) in propylene carbonate $C_4H_6O_3$ (99%, Alfa Aesar). A small part of the as-grown crystal with sizes about 5 × 5 × 1 mm (mass ~ 10–30 mg) was cleaved and mechanically fixed at the end of one of the electrodes designated as the working electrode (WE). A negative voltage $V_{CE}$ was applied to the counter electrode (CE) to extract Na$^+$ ions from the crystal. The current $I_{WE}$ (t) through the WE was used to calculate the total charge that passed through the cell and the amount of de-intercalated Na ions. The voltage $V_{RE}$ was measured by the platinum reference electrode (RE) versus WE. When the voltage source was disconnected (switch SW is off – see Fig. 2), the open circuit potential on RE designated as $V_{OCP}$ characterized the chemical potential of the sodium ions on the surface of the Na$_x$CoO$_2$ crystal attached to the WE [30]. An in-house constructed digital potentiostat designed along these lines allowed us to control and measure the voltages and currents in the cell.

The open circuit potential (OCP) measured for the as-grown Na$_x$CoO$_2$ crystal with $x \sim 0.8$ was 0.4 V. Our OCP value differed from the values reported by Berthelot et al. [22] who used a Na electrode as the reference.

To obtain crystals with a homogeneous sodium distribution, sodium extraction was conducted in small steps and equilibration was achieved in each step. First, voltage pulses $V_{CE}$ were applied in order to keep $V_{RE} = V_{target}$ constant for 2 min (inset in Fig. 3). Each voltage pulse was followed by 8 min of the open-circuit regime to achieve a homogeneous sodium distribution in the crystal. At the end of the open-circuit regime, the voltage $V_{OCP}$ on RE was recorded. During the voltage pulses $V_{CE}$, the time variation in the current $I_{WE}$ through the WE was measured and the total charge that passed through the cell was calculated.

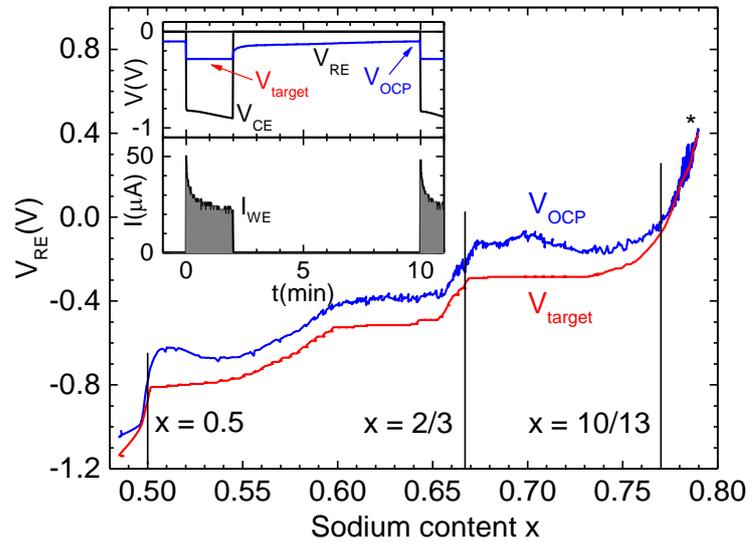

Fig. 3. Discharge curve for the as-grown crystal with an initial Na content of $x \sim 0.8$. $V_{target}$ is the target voltage on RE (red curve) and $V_{OCP}$ is the voltage at the end of the relaxation process (blue curve). Vertical lines show the positions with stable compositions. The time dependencies of the voltage and current through the electrochemical cell are shown in the inset. All voltages were measured with respect to WE. An asterisk denotes the starting point of the discharge process.

In order to vary the Na content on a large portion of the phase diagram, the voltage $V_{target}$ was scanned from 0.4 V to –1.2 V with steps of 5 mV. The position of the (*008*) XRD peak was used to confirm the sodium concentration and the sample homogeneity at the end of the electrochemical treatment.

Fig. 3 shows a typical discharge curve for the sodium cobaltate single crystals. Similar curves were obtained for various as-grown crystals with $x \sim 0.8$. At the beginning of the electrochemical treatment, the extraction of Na caused a sharp decrease in $V_{OCP}$. The sodium concentration in the $Na_xCoO_2$ crystal was estimated from the charge that passed through WE and it reached $x = 0.77$ at $V_{OCP} = -0.04$ V. This curve also exhibited two sharp drops in $V_{OCP}$. According to XRD, these $V_{OCP}$ drops corresponded to the stable sodium ordered phases of $Na_xCoO_2$, where $V_{OCP} = -0.21$ V corresponded to the phase with $x = 2/3$ and the second with $V_{OCP} = -0.83$ V corresponded to $x = 1/2$. This difference between the voltages agrees with previous electrochemical studies of polycrystalline sodium cobaltates [22]. The voltage drop associated with the $x = 0.77$ phase ($V_{OCP} = -0.04$ V) was less pronounced, which indicates that it corresponds to a smaller energy gain associated to sodium ordering. According to NMR and nuclear quadrupole resonance (NQR) data [13,23], ordered phases of $Na_xCoO_2$ for $x = 0.71$ and $0.72$ were also present, but their associated voltage drops could not be seen in the discharge curve. Therefore, the energy gained by sodium ordering with these compositions was significantly smaller [22]. The voltage plateaus on the discharge curve are the signature of solid solution behavior or of the coexistence of two phases with different sodium contents [22].

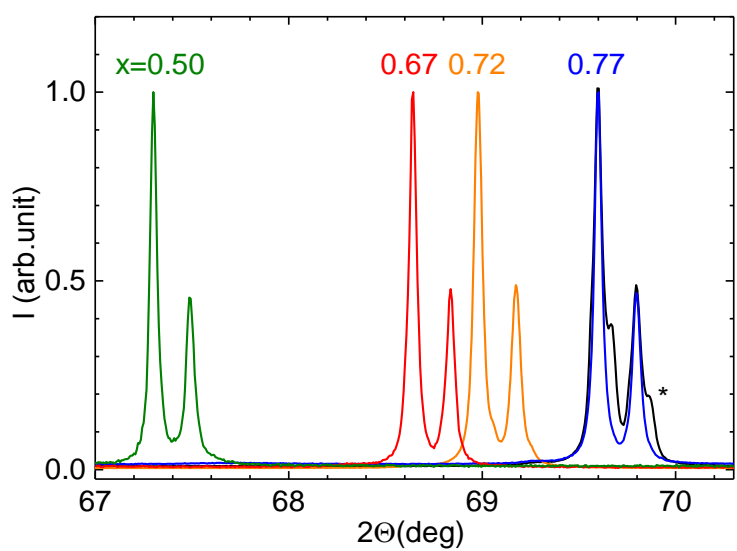

Fig. 4. XRD peaks obtained for the as-grown crystal (black) and the crystals after the electrochemical de-intercalation of Na. These data correspond to the single phase compositions obtained by progressively lowering the Na content. The secondary 0.78 phase in the as-grown crystal is denoted by an asterisk.

The voltages $V_{OCP}$ associated with voltage drops were used to reliably obtain the corresponding pure phases in a reproducible manner. The phases where $x = 0.5$, 0.67, and 0.77 were obtained by holding a constant voltage on RE or by stopping the voltage scan at the target voltage. The 0.72 phase was obtained by controlling the duration of the electrochemical treatment and the charge that passed through the cell. The (*008*) XRD peaks obtained for the as-grown two-phase crystal and crystals after electrochemical treatment are shown in Fig. 4. All the crystals yielded a narrow (*008*) peak without traces of neighboring phases.

Each full discharge curve was obtained on unique samples, but independent electrochemically synthesized crystals in the different phases were used to determine the magnetic data and they were then reduced in size for the transport measurements.

The as-grown crystals were kept in a moisture-free atmosphere in an argon-filled glove box. We found that the use of a desiccator containing silica gel as a desiccant was not suitable. Sample degradation also depended on the sample size, so the samples prepared for transport measurements were immediately inserted in the cryostat and cooled down. The storage time at room temperature in the air was reduced as much as possible. In addition, the x = 0.77 phase was less stable than the 0.5 and 0.67 phases.

**Magnetic properties**

Measuring the (*008*) peak positions by XRD allowed us to determine the sodium content and confirm the absence of secondary phases, but the XRD diffraction patterns only provides information about the thin slice of the samples which is probed by the X-rays. By contrast, magnetic susceptibility does provide information about the bulk physical properties of the samples. A Physical Properties Measurement System (PPMS) from Quantum Design equipped with a vibrating sample magnetometer was used to obtain measurements of the magnetic properties for the synthesized single crystals.

The temperature dependences of the magnetic susceptibility of the $Na_xCoO_2$ single crystals for $x = 0.5$ and 0.67 are shown in Fig. 5(a) for $H\perp c$. The data in Fig. 5(a) are compared with the magnetic susceptibility measurements obtained using a superconducting quantum interference device (SQUID) [12,23] for non-oriented polycrystalline samples with similar sodium contents. The data obtained for the powder samples and crystals were very similar. The $x = 0.67$ crystal exhibited a paramagnetic dependence of the magnetic susceptibility down to the lowest temperatures, whereas the magnetic susceptibility of the $x = 0.5$ single crystal sample depended only weakly on the temperature for $H\perp c$.

Fig. 5(b) shows the temperature dependence of the magnetic susceptibility for the $Na_{0.77}CoO_2$ single crystal and the non-oriented powder sample [28] at low temperatures. A sharp magnetic transition at $T_N = 22$ K was clearly observed for both samples, which confirmed that the same Na order occurred in both sample forms. In the AF state, the magnetic susceptibility of the $Na_{0.77}CoO_2$ single crystal was distinct from that of the powder samples, which was expected because this macroscopic response is strongly linked to the domain structure and the orientation with respect to the applied field.

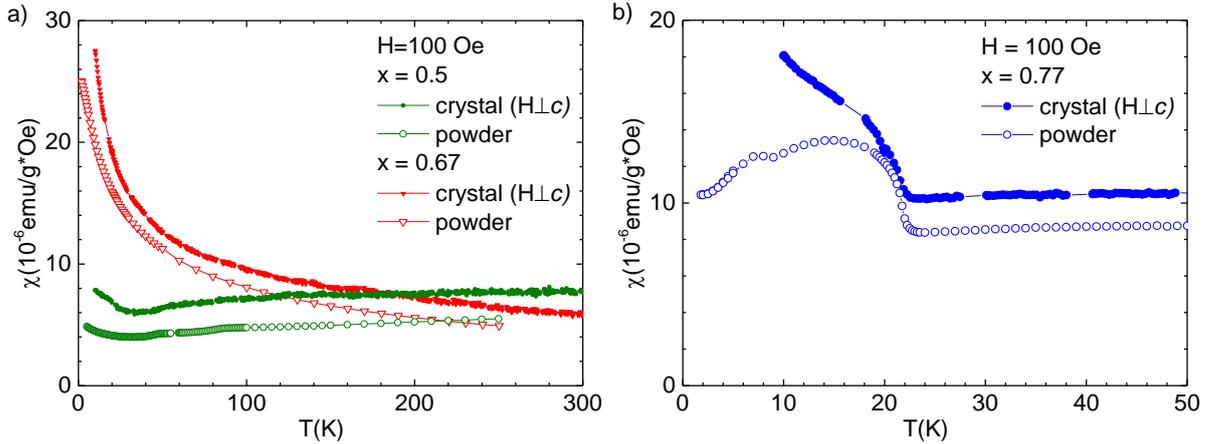

Fig. 5. Magnetic susceptibility data versus temperature ($T$) for three phases of sodium cobaltate synthesized single crystals ($H\perp c$) and non-oriented powders with the same compositions. Left panel: the large low $T$ paramagnetism is shown for the $x = 0.67$ phase. Right panel: the AF ordering at $T_N = 22$ K is shown clearly for $x = 0.77$.

In these powders, the phase purity of the samples were established from NMR data and the magnetic susceptibility data also confirmed the single phase content of the bulk single crystals. We also found that the temperature dependence of the magnetic susceptibilities of $Co_3O_4$ and $CoO$ exhibited broad peaks, which were centered at magnetic transition temperatures of $T_N = 35$ K and 290 K, respectively [16]. The absence of these peaks in Fig. 5 clearly confirmed the absence of cobalt oxide impurities in the synthesized crystals.

**Transport properties**

The resistivity of the single crystals was measured in the *ab* plane using the van der Pauw technique [31]. Square crystals with an average size $2 \times 2 \times 0.1$ mm were prepared and gold wires were attached to the samples using silver paint. Fig. 6(b) shows a photograph of a sample with the wires attached. The alternating current (AC) transport option (ACT) in PPMS was used to obtain measurements in the temperature range of 2–300 K. Resistivity measurements at temperatures of 0.1–3 K were obtained in a dilution refrigerator (BF-LD400, BlueFors Cryogenics, Finland) using the same ACT electronics. The frequency of the AC resistivity measurements was 69 Hz and the current magnitude was less than 0.5 mA at temperatures above 2 K and 0.1 mA for lower temperatures.

The temperature dependences of the in-plane resistivity for three independently synthesized $Na_{0.67}CoO_2$ crystals are shown in Fig. 6. The resistivity values for the samples were very similar and they overlapped throughout all of the temperature range. The temperature dependence of the first derivative of these resistivity curves is also shown in the figure to facilitate comparisons of these data. The good reproducibility of the residual resistivity and d$\rho$/d$T$ confirmed the high quality of the single crystal samples.

The resistivity of one $x = 0.67$ phase crystal was also measured at low temperatures and the results are shown in the inset in Fig. 6. The residual resistivity $\rho_0$ at $T = 0$ K is usually associated with scattering due to defects. We found that for our sample, $\rho_0 = 2$ μOhm·cm was much smaller than that obtained in previous studies [21,25]. The extremely low residual resistivity $\rho_0$ of our $Na_{0.67}CoO_2$ crystals confirmed their good stoichiometry.

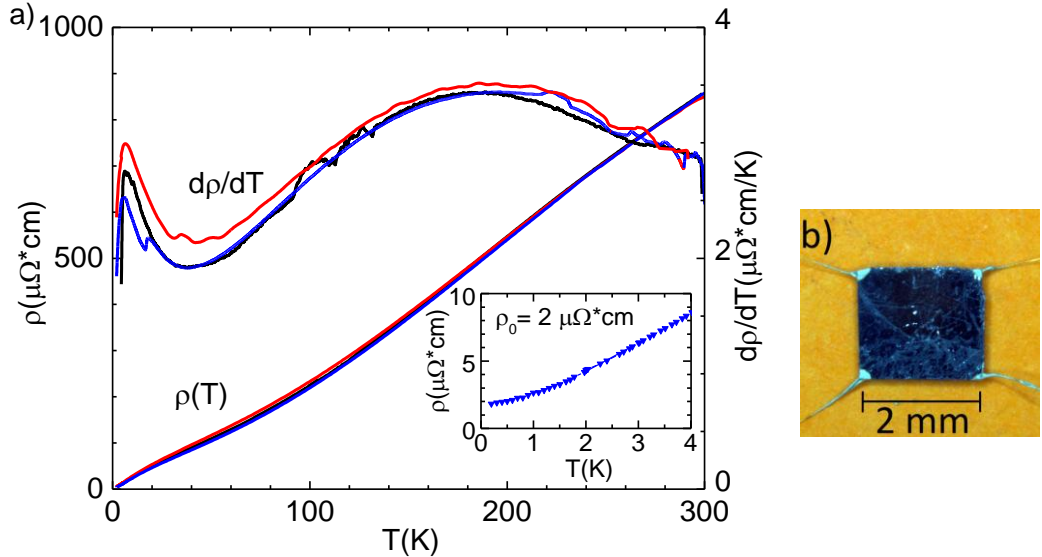

Fig. 6. Temperature dependence of the *ab* plane resistivity (left axis) and the first derivative of resistivity (right axis) for three independently synthesized $Na_{0.67}CoO_2$ crystals. Blue and red curves: samples measured in Kazan with ac current; red curve –sample measured in Saclay with dc current. An enlargement of the low temperature region is shown in the inset. Right panel: photograph of a single crystal sample cleaved in the *ab*-plane direction and prepared for resistivity measurements.

After confirming the reproducibility and reliability of our sample synthesis technique, we compared the transport properties of $Na_xCoO_2$ for three single crystal samples for $x = 0.5$, 0.67, and 0.77, which represented the three ordered phases that exhibited the most distinct behaviors of magnetic properties.

As shown in Fig. 7, these phases had quite distinct transport properties. As we shall see below, analogous resistivity curves have been also obtained in previous experiments using samples for which the Na order and the homogeneity of its content was not characterized experimentally.

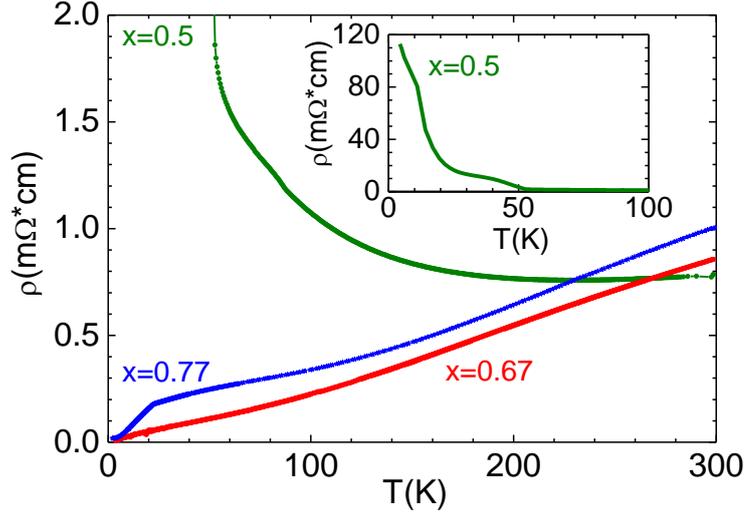

Fig. 7. Temperature dependence of the *ab* plane resistivity for $Na_xCoO_2$ where $x$ = 0.5–0.77.

The $x$ = 0.67 and 0.77 phases exhibited metallic behavior with an upward curvature of $\rho$(T) and similar $d\rho/dT$ slopes at high temperatures $T > 100$ K. However, at lower temperatures, the resistivity of the $x$ = 2/3 sample dropped drastically, as shown in Fig. 7, but the temperature variation for the $x$ = 0.77 sample flattened out and it exhibited a large drop below the Néel temperature $T_N$ = 22 K.

The behavior for $x$ = 0.67 was similar to the data reported for samples labeled as $x$ = 0.68 and 0.71 in a previous study [21], although both samples had much larger residual resistivity values than that found in the present study. For our $x$ = 2/3 sample the power law dependence of $\rho$ may correspond to the low temperature Fermi liquid type of variation, as suggested in a previous study [32] based on data obtained for specific samples [21].

For the $x$ = 0.77 magnetic phase similar behavior was found in a previous study [21] for a sample labeled as $x$ = 0.75. However, in that sample, $\rho$ (50 K) was about twice as high and the decline in the resistivity appeared to be smoother, and it did not occur near $T_N$ = 22 K.

For the $Na_{0.5}CoO_2$ phase, where the atomic structure and Na ordering have been studied previously, we found that the $\rho$ (T) curve agreed quite well with that reported in previous studies [21,24]. We note that $\rho$ (300 K) had a similar value to that found for the other phases. The resistivity increased monotonically by about a factor of two as the temperature decreased from 300 K to 53 K and a sharp phase transition occurred at 53 K, as reported previously [21,24], with another at about 20 K. Upon the establishment of the magnetically ordered states, $\rho$ (T) increased for $x$ = ½, whereas it decreased below $T_N$ = 22 K for $x$ = 0.77. In the former case, nesting of the Co bands is expected to drive the gap opening at the metal–insulator transition [33]. In the second case, the decrease in $\rho$ (T) below $T_N$ could suggest Fermi surface reconstruction in the metallic magnetic state.

**Conclusion**

In this study, we reproducibly synthesized $Na_xCoO_2$ samples with a controlled sodium concentration *x*, which corresponded to the single phases differentiated in NMR/NQR experiments on powder samples. This has been ensured by the good correspondence between the magnetic properties of our single crystals and the SQUID and NMR data obtained on oriented powder samples. Resistivity measurements taken on our well characterized samples demonstrated that the behavior was similar to that for samples synthesized with approximately similar concentrations of Na. In most cases, our samples exhibited lower residual resistivity and sharper magnetic transitions.

The *x* = 0.5, 0.67, and 0.77 phases of $Na_xCoO_2$ with specific physical properties were synthesized in a reproducible manner. Other stable sodium phases were found but they did not exhibit very sharp changes in voltage during sodium de-intercalation. Apparently, more effort will be required to obtain these phases without impurities and phase segregation. In principle, we do not envisage that any obstacle will prevent them from being obtained by applying smaller voltage steps and increasing the electrochemical treatment time. Controlling the sodium content using XRD measurements is an important procedure for checking the sample homogeneity.

Previous studies have found that it is difficult to clarify the electronic properties of these $Na_xCoO_2$ compounds as the link between magnetic, transport properties, and Na order was not established. Local-density approximations (LDA) calculations have predicted a hole band and electronic pockets, but only the former has been detected so far in angle-resolved photoemission spectroscopy (ARPES) experiments [34–36]. According to NMR experiments, we expect that the Na order and concomitant Co charge order should govern the actual band structure, as found by ARPES for some misfit cobaltates [37].

However, ARPES measurements should not significantly improve Fermi surface determinations because cleavage of the sample probably disrupts the Na order and concomitant Co charge order on the sample surface. However, Shubnikov de Haas quantum oscillations have been observed [24,25] for less characterized samples. Similar experiments to be taken on our samples should help to determine in which phases the charge order revealed by NMR induces reconstructions of the Fermi surface leading to the emergence of multiple bands at the Fermi level. This would allow us to determine the phases for which a single band model is applicable and give a simple conductivity value of $\sigma = ne^2\tau/m^*$. In cases where multiple bands must be considered at the Fermi level, we expect that detailed measurements of the Hall effect and magnetoresistance on our Na ordered samples as well as quantum oscillations data will help to better understand the ground state electronic properties. These results could be important for extending calculations [38] toward Fermi surface determinations for the ordered Na phases, including the electronic correlations and Hund coupling.

**Acknowledgments**

We thank A.V. Dooglav for help with preparing the manuscript. The crystal growth, XRD, magnetic, and transport measurements were obtained at the Federal Center of Shared Facilities of Kazan Federal University. This study was partially funded by a subsidy allocated to Kazan Federal University for state assignments in the sphere of scientific activities (project No. 3.8138.2017/8.9). G.I.F. acknowledges Russia's UMNIK FASIE in the STF program for supporting the digital potentiostat development. M.I.R. and G.I.F. are grateful for support for exchange visits to Orsay/Saclay by "Investissements d'Avenir" LabEx PALM (ANR-10-LABX-0039-PALM).